\PassOptionsToPackage{usenames,dvipsnames}{color}
\documentclass[a4paper,UKenglish]{lipics-v2016}

\usepackage{bigfoot}
\usepackage{microtype}%if unwanted, comment out or use option "draft"
\usepackage{tikz}
\usetikzlibrary{shapes.multipart, fit}
\usepackage{mathpartir}
\usepackage{stmaryrd}

\makeatletter
\newsavebox{\@brx}
\newcommand{\llangle}[1][]{\savebox{\@brx}{\(\m@th{#1\langle}\)}%
  \mathopen{\copy\@brx\kern-0.5\wd\@brx\usebox{\@brx}}}
\newcommand{\rrangle}[1][]{\savebox{\@brx}{\(\m@th{#1\rangle}\)}%
  \mathclose{\copy\@brx\kern-0.5\wd\@brx\usebox{\@brx}}}
\makeatother

\definecolor{orange}{RGB}{237,27,35}
\definecolor{blue}{RGB}{0, 83, 143}
\definecolor{green}{RGB}{247,146,29} % NOTE(dbp 2017-03-21): This is actually orange, not
\newcommand{\stlc}[1]{\mathbf{\color{Orange}{#1}}}
\newcommand{\lamref}[1]{\mathbf{\mathsf{\color{blue}{#1}}}}
\newcommand{\target}[1]{\mathtt{\color{Cyan}{#1}}}
\newcommand{\gen}[1]{\mathtt{\color{black}{#1}}}
\newcommand{\link}[1]{\mathbf{\color{Magenta}{#1}}}
\newcommand{\vo}{\mathrel{\color{black}{\vert}}}

\newcommand{\etal}{\textit{et al.}}

\title{Linking Types for Multi-Language Software:\hspace{4ex}
 Have Your Cake and Eat It Too}
\titlerunning{Linking Types for Multi-Language Software} %optional, in case that the title is too long; the running title should fit into the top page column

\author[1]{Daniel Patterson}
\author[2]{Amal Ahmed}
\affil[1]{Northeastern University, Boston MA, USA\\
  \texttt{dbp@ccs.neu.edu}}
\affil[2]{Northeastern University, Boston MA, USA\\
  \texttt{amal@ccs.neu.edu}}
\authorrunning{D. Patterson and A. Ahmed} %mandatory. First: Use abbreviated first/middle names. Second (only in severe cases): Use first author plus 'et. al.'

\Copyright{Daniel Patterson and Amal Ahmed}%mandatory, please use full first names. LIPIcs license is "CC-BY";  http://creativecommons.org/licenses/by/3.0/

\subjclass{F.3.1 Specifying and Verifying and Reasoning about Programs}% mandatory: Please choose ACM 1998 classifications from http://www.acm.org/about/class/ccs98-html . E.g., cite as "F.1.1 Models of Computation". 
\keywords{Linking, program reasoning, equivalence, expressive power of languages, fully abstract compilation}% mandatory: Please provide 1-5 keywords
\EventEditors{Benjamin S. Lerner, Rastislav Bodík, and Shriram Krishnamurthi}
\EventNoEds{2}
\EventLongTitle{2nd Summit on Advances in Programming Languages (SNAPL 2017)}
\EventShortTitle{SNAPL 2017}
\EventAcronym{SNAPL}
\EventYear{2017}
\EventDate{}
\EventLocation{}
\EventLogo{}
\SeriesVolume{}
\ArticleNo{12}

\begin{document}
\maketitle

\begin{abstract}
  Software developers compose systems from components written in many different
  languages. A business-logic component may be written in Java or OCaml, a
  resource-intensive component in C or Rust, and a high-assurance component in
  Coq. In this multi-language world, program execution sends values from one
  linguistic context to another. This boundary-crossing exposes values to
  contexts with unforeseen behavior---that is, behavior that could not arise in
  the source language of the value. For example, a Rust function may end up
  being applied in an ML context that violates the memory usage policy enforced
  by Rust's type system. This leads to the question of how developers ought to
  reason about code in such a multi-language world where behavior inexpressible
  in one language is easily realized in another.

  This paper proposes the novel idea of \emph{linking types} to address the
  problem of reasoning about single-language components in a multi-lingual
  setting. Specifically, linking types allow programmers to annotate where in a
  program they can link with components inexpressible in their unadulterated
  language. This enables developers to reason about (behavioral) equality using
  only their own language and the annotations, even though their code may be
  linked with code written in a language with more expressive power.

\vspace{0.1cm}
\noindent\emph{NOTE: This paper will be much easier to follow if viewed/printed
  in color.}

\end{abstract}

\section{Reasoning in a Multi-Language World}
\label{sec:intro}

When building large-scale software systems, programmers should be able to use
the best language for each part of the system. Using the ``best language'' means
the language that makes it is easiest for a programmer to \emph{reason} about
the behavior of that part of the system. Moreover, programmers should be able to reason
\emph{only} in that language when working on that component. That might be Rust
for a high-performance component, a terminating domain-specific language for a
protocol parser, or a general-purpose scripting language for UI code.
In some development shops, domain-specific languages are used in various parts of
systems to better separate the logic of particular problems from the plumbing
of general-purpose programming. 
But it's a
myth that programmers can reason in a single language when dealing with
multi-language software. Even if a high-assurance component is written in Coq,
the programmer must reason about extraction, compilation, and any linking that
happens at the machine-level. An ML component in a multi-language system has
contexts that may include features that don't exist in ML. This is a problem for
programmers, because as they evolve complex systems, much time is spent
refactoring---that is, making changes to components that should result in
equivalent behavior. Programmers reason about that equivalence by thinking about
possible program contexts within which the original and refactored components
could be run, though usually they only think about contexts written in their own
language. But if they have linked with another language, the additional contexts
from that language also need to be taken into account. Equivalence in all
contexts, or \emph{contextual equivalence}, is therefore central to programmer
reasoning. Unfortunately, programmers cannot rely upon contextual equivalence of
their own language. Instead, since languages interact after having been compiled
to a common target, the contextual equivalence that programmers must rely upon
is that of the compilation target, which may have little to do with their source
language.

For programmers writing components in safe languages like OCaml, the situation
is made worse by the fact that the common target is likely a low-level unsafe
language like assembly which permits direct access to memory and the call-stack. An
object-code linker will verify symbols, but little more. This means that
whenever an OCaml programmer links with C code via the FFI, they have to contend
with the fact that the C code they write can easily disrupt the equivalences
they rely on when reasoning about their OCaml code. Rather than being able to
rely upon tooling, the user of a C library must reason carefully about how the C
code will interact, at the assembly level, with their OCaml abstractions. For
example, an OCaml function that is polymorphic in its arguments could have these
arguments inspected by C code it linked with, violating parametricity. On the
other side, the C programmer attempting to write a library that can be linked
with OCaml must keep all the invariants of OCaml in mind and attempt not to
violate any of them. This is difficult and requires reasoning not only about how
the OCaml and C languages work but also how they are compiled to assembly,
because it is at the assembly level that they interact.

Since programmers use a language for its features and linguistic abstractions,
we would like programmers to be able to reason using contextual equivalence for
that language, even in the presence of target-level linking. A \emph{fully
  abstract} compiler enables exactly this reasoning: it guarantees that if two
components are contextually equivalent at the source their compiled versions are
contextually equivalent at the target. However, this guarantee comes at a steep
cost: a fully abstract compiler must disallow linking with components whose
behavior is inexpressible in the compiler's source language. But often that
extra behavior or control is exactly why the programmer is linking with a
component written in another ``more expressive'' language. An example of additional
\emph{behavior} is a non-concurrent language linking with a thread implementation
written in C. An example of additional \emph{control} is an unrestricted language
linking with a concurrent data structure written in Rust, where linear types
ensure data-race freedom.

There are two ways in which a programming language $\mathcal{A}$ can
be \emph{more expressive} than another language $\mathcal{B}$, where, 
following Felleisen~\cite{felleisen90}, we assume both languages have
been translated to a common substrate (for us, compiled to a common
target), such that $\mathcal{A}$ contexts can be wrapped around
$\mathcal{B}$ program fragments:

\begin{enumerate}
\item $\mathcal{A}$ has features unavailable in $\mathcal{B}$ that can be used
  to create contexts that can distinguish components that are contextually
  equivalent in $\mathcal{B}$. We say that language $\mathcal{A}$ is
  \emph{positively} more expressive than language $\mathcal{B}$, since the
  (larger) set of $\mathcal{A}$ contexts have more power to distinguish. For
  instance, $\mathcal{A}$ may have references or first-class control while
  $\mathcal{B}$ does not.
\item $\mathcal{A}$ has rich type-system features unavailable in $\mathcal{B}$
  that can be used to rule out contexts that, at less precise types, were able
  to distinguish inequivalent $\mathcal{B}$ components.  We say that language
  $\mathcal{A}$ is \emph{negatively} more expressive than language
  $\mathcal{B}$, since type restrictions on $\mathcal{A}$ contexts result in a
  (smaller) set of well-typed $\mathcal{A}$ contexts that have less power to
  distinguish. For instance, $\mathcal{A}$ may have linear types or polymorphism
  while $\mathcal{B}$ does not.\footnote{
    \begin{minipage}[t]{0.46\textwidth}
      Example 1: $\mathcal{A}$ has \textbf{linear types}. \\
      Consider $\mathcal{B}$ components of type $\mathtt{unit \rightarrow int}$:
      \[
      \begin{array}{l}
        \mathtt{c1}~\mbox{increments a counter on each call and returns it}\\
        \mathtt{c2}~\mbox{increments a counter on first call and returns it}
        \end{array}
      \]
      And $\mathcal{B}$ distinguishing context:
    \[
\mathtt{\lambda c.\, c\, (); c\, () : (unit \rightarrow int) \rightarrow int}     \]

      With $\mathcal{A}$ contexts of type $\mathtt{(unit
        \rightarrow int)^L \rightarrow int}$, $\mathtt{c1}$ and $\mathtt{c2}$
      are contextually equivalent, since each must be called exactly once.

\end{minipage}
\begin{minipage}[t]{0.03\textwidth}~\end{minipage}
\begin{minipage}[t]{0.48\textwidth}
  Example 2: $\mathcal{A}$ has \textbf{polymorphism}.  \\
  Consider $\mathcal{B}$ components of type $\mathtt{unit \rightarrow int}$:
\[
  \begin{array}{l}
    \mathtt{p1 = \lambda x.\, 0}\\
    \mathtt{p1 = \lambda x.\, 1}\\
  \end{array}
\]

And $\mathcal{B}$ distinguishing context:
\[
 \mathtt{\lambda f.\, if\: f()\mbox{\texttt{=}}0\: then\: diverge\: else\: () : (unit \rightarrow int) \rightarrow unit}
\]

With $\mathcal{A}$ contexts of type $\mathtt{\forall \alpha. (unit
  \rightarrow \alpha) \rightarrow unit}$, $\mathtt{p1}$ and $\mathtt{p2}$ are
contextually equivalent, since their return values may not be inspected.
\end{minipage}
    }
\end{enumerate}

The greater expressivity of programming languages explored by
Felleisen~\cite{felleisen90} is what we call \emph{positive} expressivity. As
far as we are aware, the notion of \emph{negative} expressivity, presented in
this \emph{dual} way, has not appeared in the literature.

Linking with code from more expressive languages affects not just
\emph{programmer} reasoning, but also the notion of equivalence used by
\emph{compiler writers} to justify correct optimizations. While there has
been a lot of recent work on verified compilers, most
assume no linking
(e.g.,~\cite{leroy06,leroy09:jar,lochbihler10,sevcik11,zhao13:ssa,kumar14}),
or linking only with code compiled from the same 
source
language~\cite{benton09,benton10,hur11,neis15,kang2016lightweight}.

One approach that does support cross-language linking is Compositional
Compcert~\cite{stewart15}, which nonetheless only allows linking with components
that satisfy CompCert's memory model. Another approach is the multi-language
style of verified compilers by Perconti and Ahmed~\cite{perconti14:fca}, which
allows linking with arbitrary target code that may be compiled from another
source language $\mathcal{R}$. This approach, which embeds both the source
$\mathcal{S}$ and target $\mathcal{T}$ into a single multi-language
$\mathcal{ST}$, means that compiler optimizations can be justified in terms of
$\mathcal{ST}$ contextual equivalence. However, as a tool for programmer
reasoning, this comes at a significant cost, as the programmer needs to
understand the full $\mathcal{ST}$ language and the compiler from $\mathcal{R}$
to $\mathcal{T}$. Moreover, the design of the multi-language fixes what linking
should and should not be permitted, a decision that affects the notion of
contextual equivalence used to reason about every component written in the
source language.

We contend that compiler writers should not get to decide what linking is
allowed, and indeed, we don't think they want to. Currently compiler writers are
forced to either ignore linking or make such arbitrary decisions because
existing source-language specifications are incomplete with respect to linking.
Instead, this should be a part of the language specification and
exposed to the programmer so that she can make fine-grained decisions about
linking, which leads to fine-grained control over what contexts she must
consider when reasoning about a particular component. Every compiler should then
be fully abstract, which means it preserves the equivalences chosen by the
programmer.

We advocate extending source-language specifications with
\emph{linking types}, which minimally enrich source-language types
and allow programmers to optionally annotate where in their programs
they can link with components that would not be expressible in their
unadulterated source language. As a specification mechanism, types are
familiar, and naturally allow us to change equivalences locally. They
fulfill our desire to allow the programmer fine-grained control, as
they appear on individual terms of the language. A linking-types
extension will also often introduce new terms (and operational
semantics) intended solely for reasoning about the additional contexts
introduced through linking. These new terms are a representative
abstraction of potentially complex new behavior from another language
that the programmer wants to link with. (This is analogous to how
Gu~\etal~\cite{gu15:deepspec} lift potentially complex behavior in a lower abstraction
layer into a simpler representation in a higher layer.)  Now if the
programmer reasons about contexts including those terms, she will have
considered the behavior of all contexts that her component may be
linked with after compilation. 

We envision that language designers will provide many different
linking-types extensions for their source languages. Programmers can
then opt to use zero or more of these extensions, depending on their
linking needs.

\section{Linking Types, Formally}
\label{sec:basic}

To formally present the basic idea of linking types, we consider a
setting with two simple source languages---see Figure~\ref{languages}
(top)---and show how to design linking types that mediate different
interactions between them. Our source languages are $\stlc\lambda$,
the simply typed lambda calculus with integer base types, and
$\lamref{\lambda^{ref}}$, which extends $\stlc\lambda$ with ML-like
mutable references.  We want type-preserving, fully abstract compilers
from these source languages 
to a common target language. That target should have a rich enough type system
so that the compiler's type translation can ensure full abstraction by using
types to rule out linking with target contexts whose behavior is inexpressible
in the source. Here we illustrate the idea with a fairly high-level target
language $\target{\lambda^{ref}_{exc}}$---see Figure~\ref{languages}
(bottom)---that includes mutable references and exceptions and has a modal type
system that can distinguish pure computations from those that either use
references or raise exceptions.\footnote{We use a modal type system here, but
  any type-and-effect system would suffice.} We include exceptions in the target
as a representative of the extra control flow often present in low-level targets
(e.g., direct jumps). An impure target computation
$\target{E^\bullet_{\tau_{exc}}\, \tau}$ (pronounced ``impure exception-raising tau
computation'') may access the heap while computing a value of type $\target\tau$
or raising an exception of type $\target{\tau_{exc}}$. In contrast, a pure
computation $\target{E^\circ_0\, \tau}$ (pronounced ``pure tau computation'') may not
access the heap, and cannot raise exceptions as the exception type is the void
(uninhabited) type $\target{0}$.

\begin{figure}[t]
  \begin{small}
    \begin{minipage}[t]{0.5\textwidth}
      \[
      \begin{array}{lrcl}
        \stlc\lambda & \stlc{\tau} & ::=
        & \stlc{unit \vo int \vo \tau \rightarrow \tau} \\
        
                    & \stlc{e} & ::=
        & \stlc{() \vo n \vo x \vo \lambda x\mathbin{\colon}\tau.\, e \vo e\, e }\\
                    & & & \stlc{e + e \vo e * e \vo e - e}\\
                    & \stlc{v} & ::=
        & \stlc{() \vo n \vo \lambda x\mathbin{\colon}\tau.\, e}\\
      \end{array}
      \]
    \end{minipage}
    \begin{minipage}[t]{0.5\textwidth}
      \[
      \begin{array}{lrcl}
        \lamref{\lambda^{ref}} & \lamref{\tau} &::=
        & \lamref{\ldots \vo ref\, \tau} \\
        
                            & \lamref{e} &::=
        & \lamref{\ldots \vo ref\, e \vo e := e \vo \mathbin{!} e} \\

                            & \lamref{v} & ::=
        & \lamref{\ldots \vo \ell}\\
      \end{array}
      \]
    \end{minipage}

    \vspace{0.25cm}
    \hrule
    \[
      \begin{array}{lrcl} 
        \target{\lambda^{ref}_{exc}} & \target{\tau} & ::=
        & \target{0 \vo unit \vo int \vo ref\, \tau \vo \tau \rightarrow E^\epsilon_{\tau_{exc}}\, \tau} \\
                      & \target{\epsilon} & ::= & \target{\bullet \vo \circ} \\
        
                      & \target{e} & ::=
        & \target{() \vo n \vo x \vo \lambda x\mathbin{\colon}\tau.\, e \vo e\, e \vo e + e \vo e * e \vo e - e \vo throw\, e}\\
                    & & & \target{catch\, e\, with\, val\, x \Rightarrow e\, ;\, exc\, y \Rightarrow e \vo ref\, e \vo e := e \vo \mathbin{!} e} \\
        & \target{v} & ::=
        & \target{() \vo n \vo \lambda x\mathbin{\colon}\tau.\, e \vo \ell}
      \end{array}      
    \]

    \fbox{$\Gamma \vdash \target v : \target\tau$}
    \vspace{-0.85cm}
    \begin{mathpar}  
      \inferrule{ }{\Gamma \vdash \target{()} \colon \target{unit}}
      \and
      \inferrule
      {\Gamma, \target{x} \colon \target\tau \vdash \target{e \colon E^\rho_{\tau_{exn}} \tau'}}
      {\Gamma \vdash \target{\lambda\, x \colon \tau . e} \colon \target{\tau \rightarrow E^\rho_{\tau_{exn}} \tau'}}
    \end{mathpar}

    \fbox{$\Gamma \vdash \target e \colon \target{E^\epsilon_{\tau_{exn}} \tau}$}
    \vspace{-0.75cm}
    \begin{mathpar}
      \hspace{2.5cm}
      \inferrule{\Gamma \vdash \target{v} \colon \target{\tau}}{\Gamma \vdash \target{v} \colon \target{E^\circ_0 \tau}}
      \quad
      \inferrule{\Gamma \vdash \target{e_1} \colon
        \target{E^{\rho_1}_{\tau_{exn}} (\tau \rightarrow E^{\rho_3}_{\tau_{exn}} \tau')} \\ \Gamma \vdash \target{e_2 \colon \target{E^{\rho_2}_{\tau_{exn}} \tau}}}{\Gamma \vdash \target{e_1\, e_2} \colon \target{E^{\rho_1\lor\rho_2\lor\rho_3}_{\tau_{exn}} \tau'}}
      \quad
      \inferrule
      {\Gamma \vdash \target{e} \colon \target{E^\rho_{\tau_{exn}}\, \tau} \\ \vdash \target\tau}
      {\Gamma \vdash \target{ref\, e \colon E^\bullet_{\tau_{exn}}\, ref \tau}}

      \inferrule
      {\Gamma \vdash \target{e_1 \colon E^{\rho_1}_{\tau_{exn}}\, ref\, \tau} \\ \Gamma \vdash \target{e_1 \colon  E^{\rho_2}_{\tau_{exn}}\, \tau}}
      {\Gamma \vdash \target{e_1 := e_2 \colon E^\bullet_{\tau_{exn}}\, unit}}

      \inferrule
      {\Gamma \vdash \target{e \colon E^\rho_{\tau_{exn}}\, ref\, \tau}}
      {\Gamma \vdash \target{!e_1 \colon E^\bullet_{\tau_{exn}}\, \tau}}

      \inferrule
      {\Gamma \vdash \target{e} \colon \target{E^\rho_{\tau_{exn}}\, \tau} \\
        \Gamma,\target{x} \colon \target\tau \vdash \target{e_2} \colon \target{E^{\rho_2}_{\tau_{exn}'}\, \tau'} \\
        \Gamma,\target{y} \colon \target{\tau_{exn}} \vdash \target{e_1} \colon \target{E^{\rho_1}_{\tau_{exn}'}\, \tau'}}
      {\Gamma \vdash \target{catch\, e\, with\, val\, x \Rightarrow e_1\, ;\, exc\, y \Rightarrow e_2} \colon \target{E^{\rho_1\lor\rho_2}_{\tau_{exn}'}\, \tau'}}
      
      \inferrule
      {\Gamma \vdash \target{e} \colon \target{\tau_{exn}} \\ \vdash \target\tau}
      {\Gamma \vdash \target{throw\, e \colon E^\circ_{\tau_{exn}}\, \tau}}

    \end{mathpar}
    \caption{$\stlc\lambda$ and $\lamref{\lambda^{ref}}$ syntax (top), $\target{\lambda^{ref}_{exc}}$ syntax and selected static semantics (bottom).}
    \label{languages}
  \end{small}
\end{figure}

Consider the scenario where the programmer writes code in
$\stlc\lambda$ and wants to link with code written in
$\lamref{\lambda^{ref}}$. Assume this linking happens after both
$\stlc\lambda$ and $\lamref{\lambda^{ref}}$ have been compiled using 
\emph{fully abstract} compilers to $\target{\lambda^{ref}_{exc}}$.
We illustrate this with concrete example programs $\stlc{e_1}$ and $\stlc{e_2}$ which are equivalent
in $\stlc\lambda$. Now consider the context $\lamref{C^{ref}}$ which implements a simple
counter using a reference cell.
The $\stlc\lambda$ compiler, since it is fully abstract, would have to
disallow linking with $\lamref{C^{ref}}$ since it can distinguish
$\stlc{e_1}$ from $\stlc{e_2}$.  More generally, in order
to rule out this class of equivalence-disrupting contexts, the fully
abstract compiler would have to prevent linking with any code that has
externally visible effects.\footnote{For simplicity, the type system
  we show here doesn't support effect masking, so we rule out linking
  with all effectful code.  More realistic target languages,
  e.g., based on Koka~\cite{leijen14}, would support linking with code without
  externally visible effects.} This can be accomplished by a
type-directed compiler that sends all $\stlc\lambda$ arrows
$\stlc{\tau_1 \rightarrow \tau_2}$ to pure functions
$\target{\tau_1' \rightarrow \target{E^\circ_0\, \tau_2'}}$, where
$\target{\tau_1'}$ and $\target{\tau_2'}$ are the translations of
types $\stlc{\tau_1}$ and $\stlc{\tau_2}$.  This would rule out
linking with contexts with heap effects like
$\lamref{C^{ref}}$. But in this case, the programmer wants to link these
together and is willing to lose some equivalences in order to do so.

\begin{small}
  \begin{minipage}{0.4\textwidth}
    \[
      \begin{array}{l}
        \stlc{e_1} = \stlc{\lambda c.\, c ()} \\
        \stlc{e_2} = \stlc{\lambda c.\, c (); c ()} \\
        \\
        \forall \stlc{C^\lambda}. \stlc{C^\lambda}[\stlc{e_1}]
        \approx_{\stlc{\lambda}} \stlc{C^\lambda}[\stlc{e_2}] \\
      \end{array}
    \]
  \end{minipage}
  \begin{minipage}{0.6\textwidth}
    \[
    \begin{array}{l}
      \lamref{C^{ref}} = \begin{array}[t]{l}
                           \lamref{let\, x = ref\, 0\, in}\\
                           \lamref{let\, c'\, () = x :=\, !x + 1; !x\, in\,
                           \gen{[\cdot]} c'}
                           \end{array}\\
      \lamref{C^{ref}}[\stlc{e_1}] \Downarrow 1 \\
      \lamref{C^{ref}}[\stlc{e_2}] \Downarrow 2 \\
    \end{array}
  \]
\end{minipage}
\end{small}

\begin{figure}[t]
  \begin{small}

    \begin{minipage}[t]{0.5\textwidth}

    \[
      \begin{array}{llcl}
        \stlc{\lambda}^{\link\kappa} & \link{\tau} & ::= & \link{unit \vo int \vo ref\, \tau \vo \tau \rightarrow R^\epsilon\, \tau}\\
        & \link{e} &::= &
                          \stlc{() \vo n \vo x \vo \lambda x\mathbin{\colon}\link\tau.\, \link{e} \vo  \link{e}\,  \link{e} \vo  \link{e} +  \link{e}} \\
        & & & \stlc{\vo \link{e} *  \link{e} \vo  \link{e} -  \link{e}} \vo \link{ref\, e \vo e := e \vo ! e} \\
        
        & \link{v} & ::= & \stlc{() \vo n \vo \lambda x\mathbin{\colon}\link\tau.\, \link{e}} \vo \link{\ell}\\
        & \link{\epsilon} & ::= & \link{\bullet \vo \circ}\\
      \end{array}
    \]
      \[
        \begin{array}{lcl}
        \link{\kappa}^+(\stlc{unit}) & = & \link{unit} \\
        \link{\kappa}^+(\stlc{int}) & = & \link{int} \\
        \link{\kappa}^+(\stlc{\tau_1 \rightarrow \tau_2}) & = & \link{\kappa}^+(\stlc{\tau_1}) \link{\rightarrow R^\circ\, \kappa}^+(\stlc{\tau_2})\\
        \end{array}
      \]
      \[
        \begin{array}{lcl}
          \link{\kappa}^-(\link{unit}) & = & \stlc{unit} \\
          \link{\kappa}^-(\link{int}) & = & \stlc{int} \\
          \link{\kappa}^-(\link{ref\, \tau}) & = & \link{\kappa}^-(\link\tau) \\
          \link{\kappa}^-(\link{\tau_1 \rightarrow R^\epsilon\, \tau_2}) & = & \link{\kappa}^-(\link{\tau_1}) \stlc\rightarrow \link{\kappa}^-(\link{\tau_2})\\
        \end{array}
      \]
    \end{minipage}
    \begin{minipage}[t]{0.5\textwidth}
      \[
      \begin{array}{llcl}
        \lamref{\lambda}^{\lamref{ref}\link\kappa} & \link{\tau} & ::= & \link{unit \vo int \vo ref\, \tau \vo \tau \rightarrow R^\epsilon\, \tau}\\
        & \link{e} &::= &
                          \lamref{() \vo n \vo x \vo \lambda x\mathbin{\colon}\link\tau.\, \link{e} \vo  \link{e}\,  \link{e} \vo  \link{e} +  \link{e}} \\
        & & & \lamref{\vo \link{e} *  \link{e} \vo  \link{e} -  \link{e} \vo ref\, e \vo e := e \vo ! e} \\
        
        & \link{v} & ::= & \lamref{() \vo n \vo \lambda x\mathbin{\colon}\link\tau.\, \link{e} \vo \ell}\\
        & \link{\epsilon} & ::= & \link{\bullet \vo \circ}\\
      \end{array}
    \]
    \[
      \begin{array}{lcl}
        \link{\kappa}^+(\lamref{unit}) & = & \link{unit} \\
        \link{\kappa}^+(\lamref{int}) & = & \link{int} \\
                \link{\kappa}^+(\lamref{ref\, \tau}) & = & \link{ref\, \kappa^+(\lamref{\tau})} \\ 
        \link{\kappa}^+(\lamref{\tau_1 \rightarrow \tau_2}) & = & \link{\kappa}^+(\lamref{\tau_1}) \link{\rightarrow R^\bullet\, \kappa}^+(\lamref{\tau_2})\\
      \end{array}
    \]
    \[
      \begin{array}{lcl}
        \link{\kappa}^-(\link{unit}) & = & \lamref{unit} \\
          \link{\kappa}^-(\link{int}) & = & \lamref{int} \\
          \link{\kappa}^-(\link{ref\, \tau}) & = & \lamref{ref}\, \link{\kappa}^-(\link\tau) \\
          
        \link{\kappa}^-(\link{\tau_1 \rightarrow R^\epsilon\, \tau_2}) & = & \link{\kappa}^-(\link{\tau_1}) \lamref\rightarrow \link{\kappa}^-(\link{\tau_2})\\
      \end{array}
    \]
  \end{minipage}

    \caption{Linking-types extension of $\stlc\lambda$ and $\lamref{\lambda^{ref}}$.}
    \label{basic-linking-types} 
  \end{small}
\end{figure}

To enable the above linking, we present a linking-types extension for
$\stlc\lambda$ that includes both an extended language
$\stlc\lambda^{\link\kappa}$ and functions $\link\kappa^+$ and $\link\kappa^-$
that relate types of $\stlc\lambda$ and $\stlc\lambda^{\link\kappa}$. The
$\stlc\lambda^{\link\kappa}$ type system includes reference types and tracks
heap effects. We need to track heap effects to be able to reason about the
interaction between the pure $\stlc\lambda$ code and impure
$\lamref{\lambda^{ref}}$ code that it will be linked with. This extension is
shown on the left in Figure~\ref{basic-linking-types}. The parts of
$\stlc\lambda^{\link\kappa}$ that extend $\stlc\lambda$ are typeset in
$\link{magenta}$, whereas terms that originated in $\stlc\lambda$ are
$\stlc{orange}$. $\stlc\lambda^{\link\kappa}$ types $\link\tau$ include base
types $\link{unit}$ and $\link{int}$, reference types $\link{ref\, \tau}$, and a
computation type $\link{R^\epsilon\,\tau}$, analogous to the target computation
type $\target{E^\epsilon_{\tau_{exn}}\tau}$, but without tracking exception
effects. $\stlc\lambda^{\link\kappa}$ terms $\link{e}$ include terms from
$\stlc\lambda$, as well as terms for allocating, reading, and updating
references.

With this extension, we annotate $\stlc{e_1}$ and $\stlc{e_2}$
with a linking type that specifies that the input can be heap-effecting: $\stlc{\lambda c.\, c ()}
\not\approx_{\stlc\lambda^{\link\kappa}}^{ctx} \stlc{\lambda c.\, c (); c ()
  \colon \link{(unit \rightarrow R^\bullet\, int) \rightarrow R^\bullet\,
    int}}$. At this type, $\stlc{e_1}$ and $\stlc{e_2}$ are no longer contextually equivalent
and, further, can be linked with the counter library.

Without the above annotation, the compiler would translate the type of
$\stlc{\lambda c.\, c ()}$ or $\stlc{\lambda c.\, c (); c ()}$ from
the $\stlc{\lambda}$ type $\stlc{unit \rightarrow int}$ to the
$\target{\lambda^{ref}_{exn}}$ type $\target{unit \rightarrow
  E^\circ_0 int}$, and the type expected by the counter from the
$\lamref{\lambda^{ref}}$ type $\lamref{unit \rightarrow int}$ to the
$\target{\lambda^{ref}_{exn}}$ type $\target{unit \rightarrow
  E^\bullet_0 int}$. Since these are not the same, an error would be
reported: that $\stlc{unit \rightarrow int}$ is not compatible with
$\lamref{unit \rightarrow int}$. This error matches our intuition ---
that an arrow means something fundamentally different in a pure
language and one that has heap effects. For 
advanced users, the compiler could explain the type translations that gave rise
to that incompatibility. By contrast, with the type annotation $\link{unit
  \rightarrow R^\bullet\, int}$ both types translate to the same
$\target{\lambda^{ref}_{exn}}$ type $\target{unit \rightarrow
  E^\bullet_0 int}$ and thus no error will be raised. 

With the linking-types-extended language, note that the additional \emph{terms}
are intended only for reasoning, so that programmers can understand the kind of
behavior that they are linking with; they should not show up in code written by
the programmer. If we allowed programmers to use these terms in their code, we
would be changing the programming language itself, whereas linking types should
only allow a programmer to change equivalences of their existing language. Our
focus is \emph{linking}, not general language extension. The last part of the
linking-types extension is the pair of functions $\link\kappa^+$, for embedding
$\stlc\lambda$ types in $\stlc\lambda^{\link\kappa}$, and $\link\kappa^-$ for
projecting $\stlc\lambda^{\link\kappa}$ types to $\stlc\lambda$ types. We will
discuss the properties that $\link\kappa^+$ and $\link\kappa^-$ must satisfy
below.

Also shown in Figure~\ref{basic-linking-types} is a linking-types
extension of $\lamref{\lambda^{ref}}$ that allows
$\lamref{\lambda^{ref}}$ to distinguish program fragments that are
free of heap effects and can then safely be passed to linked
$\stlc\lambda$ code. This results in essentially the same extended
language $\stlc\lambda^{\link\kappa}$; the only changes are the arrow
and reference cases of $\link{\kappa}^+$ and $\link{\kappa}^-$ and in
terms that should be written by programmers.

We can now develop fully abstract compilers from
$\stlc{\lambda^{\link\kappa}}$ and
$\lamref{\lambda^{ref\link\kappa}}$---rather than $\stlc{\lambda}$ and
$\lamref{\lambda^{ref}}$---to $\target{\lambda^{ref}_{exc}}$ using the
following type translation to ensure full abstraction:
\[
  \begin{array}{ll}
    \llangle \link{unit} \rrangle & = \target{unit} \\[0.25em]
    \llangle \link{int} \rrangle & = \target{int} \\[0.25em]
    \llangle \link{ref\, \tau} \rrangle & = \target{ref\, \llangle \link{\tau} \rrangle} \\[0.25em]
    \llangle \link{\tau_1 \rightarrow R^\epsilon\, \tau_2} \rrangle & = \llangle
    \link{\tau_1} \rrangle \target{\rightarrow} \target{E_0^\epsilon\, \llangle
    \link{\tau_2} \rrangle} \\
    \end{array}
  \]

\subsection{Properties of Linking Types}
For any source language $\gen{\lambda_{src}}$, an extended language
$\gen{\lambda_{src}^{\link{\kappa}}}$ paired with $\link\kappa^+$ and
$\link\kappa^-$ is a linking-types extension if the following
properties hold:

\begin{itemize}
\item $\gen{\lambda_{src}}$ terms are a subset of
  $\gen\lambda_{\gen{src}}^{\link\kappa}$ terms.  
\item $\gen{\lambda_{src}}$ type $\gen\tau$ embeds into a $\gen\lambda_{\gen{src}}^{\link\kappa}$ type by $\link{\kappa}^+(\gen{\tau})$.
\item $\gen\lambda_{\gen{src}}^{\link\kappa}$ type
  $\link{\tau^\kappa}$ projects to a $\gen{\lambda_{src}}$ type by
  $\link{\kappa}^-(\link{\tau^\kappa})$.
\item For any $\gen{\lambda_{src}}$ type $\gen\tau$, $\link{\kappa}^-(\link{\kappa}^+(\gen\tau)) = \gen\tau$.
\item $\link\kappa^+$ preserves and reflects equivalence: \\
  $\forall \gen{e_1},\gen{e_2} \in \gen\lambda_{\gen{src}}.~ \gen{e_1} \approx^{ctx}_{\lambda_{\gen{src}}} \gen{e_2} : \tau
  \Longleftrightarrow \gen{e_1} \approx^{ctx}_{\lambda_{\gen{src}}^{\link\kappa}} \gen{e_2} :
  \link\kappa^+(\tau)$.
\item
  $\forall \gen{e}, \link{\tau}.~ \gen{e} : \link{\tau} \implies
  \gen{e} : \link{\kappa}^-(\link{\tau})$ when $\gen{e}$ only
  contains $\lambda_{\gen{src}}$ terms.
\item A compiler for $\gen\lambda_{\gen{src}}^{\link\kappa}$ should be
  fully abstract, but it need only compile terms from $\gen{\lambda_{src}}$.
\end{itemize}

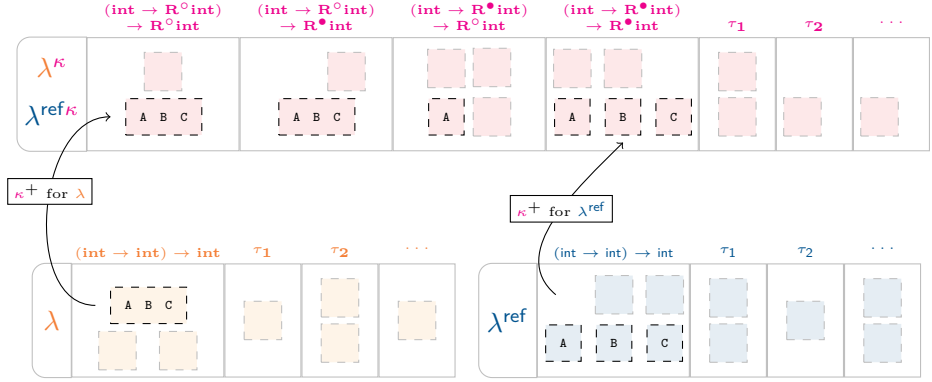
\begin{figure}
  \[
    \begin{array}{ll}
      \gen{program\, A} & \gen{\lambda f:int \rightarrow int.\, 1} \\
      \gen{program\, B} & \gen{\lambda f:int \rightarrow int.\, f\, 0;\, 1} \\
      \gen{program\, C} & \gen{\lambda f:int \rightarrow int.\, f\, 0;\, f\, 0;\, 1} \\
    \end{array}
  \]

\pgfdeclarelayer{background}
\pgfsetlayers{background,main}

\begin{tikzpicture}
  \begin{tiny}
    \begin{pgfonlayer}{main} 
      \draw[gray!50] node[minimum width=0.9cm, minimum height=1.5cm,
      append after command={[rounded corners=5pt](left.west)|-(left.north)},
      append after command={[rounded corners=0pt](left.north)-|(left.east)},
      append after command={[rounded corners=0pt](left.east)|-(left.south)},
      append after command={[rounded corners=5pt](left.south)-|(left.west)}] 
      (left) {};
      
      \node[draw=gray!50, minimum height=1.5cm, minimum width=2cm, anchor=west] (pp) at (left.east) {};
      \node [anchor=north, align=center] at ([shift={(0,0.55cm)}]pp.north) {$\link{(int \rightarrow R^\circ int)}$\\ $\link{\rightarrow R^\circ int}$};

      \node[draw=gray!50, minimum height=1.5cm, minimum width=2cm, anchor=west] (pi) at (pp.east) {};
      \node [anchor=north, align=center] at ([shift={(0,0.55cm)}]pi.north) {$\link{(int \rightarrow R^\circ int)}$\\ $\link{\rightarrow R^\bullet int}$};

      \node[draw=gray!50, minimum height=1.5cm, minimum width=2cm, anchor=west] (ip) at (pi.east) {};
      \node [anchor=north, align=center] at ([shift={(0,0.55cm)}]ip.north) {$\link{(int \rightarrow R^\bullet int)}$\\ $\link{\rightarrow R^\circ int}$};

      \node[draw=gray!50, minimum height=1.5cm, minimum width=2cm, anchor=west] (ii) at (ip.east) {};
      \node [anchor=north, align=center] at ([shift={(0,0.55cm)}]ii.north) {$\link{(int \rightarrow R^\bullet int)}$\\ $\link{\rightarrow R^\bullet int}$};

      \node[draw=gray!50, minimum height=1.5cm, minimum width=1cm, anchor=west] (kmisc1) at (ii.east) {};
      \node [anchor=north, align=center] at ([shift={(0,0.3cm)}]kmisc1.north) {$\link{\tau_1}$};
      
      \node[draw=gray!50, minimum height=1.5cm, minimum width=1cm, anchor=west] (kmisc2) at (kmisc1.east) {};
      \node [anchor=north, align=center] at ([shift={(0,0.3cm)}]kmisc2.north) {$\link{\tau_2}$};

      \draw node[draw=gray!50, minimum width=1cm, minimum height=1.5cm, anchor=west]
      (right) at (kmisc2.east) {};
      \node [anchor=north, align=center] at ([shift={(0,0.3cm)}]right.north) {$\link{\ldots}$};

      \node[] at ([shift={(0,0.35cm)}]left.center) {\large{$\stlc{\lambda^{\link\kappa}}$}};
      \node[] at ([shift={(0,-0.25cm)}]left.center) {\large{$\lamref{\lambda^{ref\link\kappa}}$}};

      \draw[gray!50] node[minimum width=0.5cm, minimum height=1.5cm,
      append after command={[rounded corners=5pt](leftp.west)|-(leftp.north)},
      append after command={[rounded corners=0pt](leftp.north)-|(leftp.east)},
      append after command={[rounded corners=0pt](leftp.east)|-(leftp.south)},
      append after command={[rounded corners=5pt](leftp.south)-|(leftp.west)}] 
      (leftp) at (0cm,-3cm) {};

      \node[] at (leftp.center) {\large{$\stlc\lambda$}}; 
      
      \node[draw=gray!50, minimum height=1.5cm, minimum width=2cm, anchor=west] (p) at (leftp.east) {};
      \node [anchor=north, align=center] at ([shift={(0,0.3cm)}]p.north) {$\stlc{(int \rightarrow int) \rightarrow int}$};

      \node[draw=gray!50, minimum height=1.5cm, minimum width=1cm, anchor=west] (p1) at (p.east) {};
      \node [anchor=north, align=center] at ([shift={(0,0.3cm)}]p1.north) {$\stlc{\tau_1}$};

       \node[draw=gray!50, minimum height=1.5cm, minimum width=1cm, anchor=west] (p2) at (p1.east) {};
      \node [anchor=north, align=center] at ([shift={(0,0.3cm)}]p2.north) {$\stlc{\tau_2}$};

       \draw node[draw=gray!50, minimum width=1cm, minimum height=1.5cm, anchor=west] 
      (rightp) at (p2.east) {};
      \node [anchor=north, align=center] at ([shift={(0,0.3cm)}]rightp.north) {$\stlc{\ldots}$};

      \draw[gray!50] node[minimum width=0.75cm, minimum height=1.5cm,
      append after command={[rounded corners=5pt](leftr.west)|-(leftr.north)},
      append after command={[rounded corners=0pt](leftr.north)-|(leftr.east)},
      append after command={[rounded corners=0pt](leftr.east)|-(leftr.south)},
      append after command={[rounded corners=5pt](leftr.south)-|(leftr.west)}] 
      (leftr) at (6cm,-3cm) {};

      \node[] at (leftr.center) {\large{$\lamref{\lambda^{ref}}$}};

      \node[draw=gray!50, minimum height=1.5cm, minimum width=2cm, anchor=west] (r1) at (leftr.east) {};
      \node [anchor=north, align=center] at ([shift={(0,0.3cm)}]r1.north) {$\lamref{(int \rightarrow int) \rightarrow int}$};

      \node[draw=gray!50, minimum height=1.5cm, minimum width=1cm, anchor=west] (r2) at (r1.east) {};
      \node [anchor=north, align=center] at ([shift={(0,0.3cm)}]r2.north) {$\lamref{\tau_1}$};

      \node[draw=gray!50, minimum height=1.5cm, minimum width=1cm, anchor=west] (r3) at (r2.east) {};
      \node [anchor=north, align=center] at ([shift={(0,0.3cm)}]r3.north) {$\lamref{\tau_2}$};
      
       \draw node[draw=gray!50, minimum width=1cm, minimum height=1.5cm, anchor=west] 
      (rightr) at (r3.east) {};
      \node [anchor=north, align=center] at ([shift={(0,0.3cm)}]rightr.north) {$\lamref{\ldots}$};

      \node[] (A) at ([shift={(0.75cm,0.2cm)}]p.west) {$\gen{A}$};
      \node[anchor=west] (B) at (A.east) {$\gen{B}$};
      \node[anchor=west] (C) at (B.east) {$\gen{C}$};
      
      \node[] (App) at ([shift={(0.75cm,-0.3cm)}]pp.west) {$\gen{A}$};
      \node[anchor=west] (Bpp) at (App.east) {$\gen{B}$};
      \node[anchor=west] (Cpp) at (Bpp.east) {$\gen{C}$};

      \node[] (Api) at ([shift={(0.75cm,-0.3cm)}]pi.west) {$\gen{A}$};
      \node[anchor=west] (Bpi) at (Api.east) {$\gen{B}$};
      \node[anchor=west] (Cpi) at (Bpi.east) {$\gen{C}$};

      \node[] (Aip) at ([shift={(-0.3cm,-0.3cm)}]ip.center) {$\gen{A}$};

      \node[] (Aii) at ([shift={(0.35cm,-0.3cm)}]ii.west) {$\gen{A}$}; 
      \node[anchor=west] (Bii) at ([shift={(0.4cm,0cm)}]Aii.east) {$\gen{B}$}; 
      \node[anchor=west] (Cii) at ([shift={(0.4cm,0cm)}]Bii.east) {$\gen{C}$};

      \node[] (Ar) at ([shift={(0.35cm,-0.3cm)}]r1.west) {$\gen{A}$}; 
      \node[anchor=west] (Br) at ([shift={(0.4cm,0cm)}]Ar.east) {$\gen{B}$};
      \node[anchor=west] (Cr) at ([shift={(0.4cm,0cm)}]Br.east) {$\gen{C}$};

      \node[text=orange!10, draw=gray!50, thin, dashed, inner sep=2mm, outer sep=2mm, fill=orange!50, fill opacity=0.2] at ([shift={(0.35cm,-0.3cm)}]kmisc2.west) {$\mathtt{X}$};
      \node[text=orange!10, draw=gray!50, thin, dashed, inner sep=2mm, outer sep=2mm, fill=orange!50, fill opacity=0.2] at ([shift={(0,0.5cm)}]Aip.north) {$\mathtt{X}$};
      \node[text=orange!10, draw=gray!50, thin, dashed, inner sep=2mm, outer sep=2mm, fill=orange!50, fill opacity=0.2] at ([shift={(0.35cm,-0.3cm)}]right.west) {$\mathtt{X}$};
      \node[text=orange!10, draw=gray!50, thin, dashed, inner sep=2mm, outer sep=2mm, fill=orange!50, fill opacity=0.2] (misc1) at ([shift={(0.3cm,-0.3cm)}]ip.center) {$\mathtt{X}$};
      \node[text=orange!10, draw=gray!50, thin, dashed, inner sep=2mm, outer sep=2mm, fill=orange!50, fill opacity=0.2] at ([shift={(0,0.2cm)}]misc1.north) {$\mathtt{X}$};
      \node[text=orange!10, draw=gray!50, thin, dashed, inner sep=2mm, outer sep=2mm, fill=orange!50, fill opacity=0.2] at ([shift={(0,-0.3cm)}]kmisc1.center) {$\mathtt{X}$};
      \node[text=orange!10, draw=gray!50, thin, dashed, inner sep=2mm, outer sep=2mm, fill=orange!50, fill opacity=0.2] at ([shift={(0,0.3cm)}]kmisc1.center) {$\mathtt{X}$};
      \node[text=orange!10, draw=gray!50, thin, dashed, inner sep=2mm, outer sep=2mm, fill=orange!50, fill opacity=0.2] at ([shift={(0,0.3cm)}]pp.center) {$\mathtt{X}$};
      \node[text=orange!10, draw=gray!50, thin, dashed, inner sep=2mm, outer sep=2mm, fill=orange!50, fill opacity=0.2] at ([shift={(0.4cm,0.3cm)}]pi.center) {$\mathtt{X}$};
      \node[text=orange!10, draw=gray!50, thin, dashed, inner sep=2mm, outer sep=2mm, fill=orange!50, fill opacity=0.2] at ([shift={(0cm,0.5cm)}]Aii.north) {$\mathtt{X}$};
      \node[text=orange!10, draw=gray!50, thin, dashed, inner sep=2mm, outer sep=2mm, fill=orange!50, fill opacity=0.2] at ([shift={(0cm,0.5cm)}]Bii.north) {$\mathtt{X}$}; 
      
      \node[text=green!10, draw=gray!50, thin, dashed, inner sep=2mm, outer sep=2mm, fill=green!50, fill opacity=0.2] at ([shift={(0.4cm,-0.4cm)}]p.center) {$\mathtt{X}$};
      \node[text=green!10, draw=gray!50, thin, dashed, inner sep=2mm, outer sep=2mm, fill=green!50, fill opacity=0.2] at ([shift={(-0.4cm,-0.4cm)}]p.center) {$\mathtt{X}$};
      \node[text=green!10, draw=gray!50, thin, dashed, inner sep=2mm, outer sep=2mm, fill=green!50, fill opacity=0.2] at (p1.center) {$\mathtt{X}$};
      \node[text=green!10, draw=gray!50, thin, dashed, inner sep=2mm, outer sep=2mm, fill=green!50, fill opacity=0.2] at ([shift={(0,-0.3cm)}]p2.center) {$\mathtt{X}$};
      \node[text=green!10, draw=gray!50, thin, dashed, inner sep=2mm, outer sep=2mm, fill=green!50, fill opacity=0.2] at ([shift={(0,0.3cm)}]p2.center) {$\mathtt{X}$};
      \node[text=green!10, draw=gray!50, thin, dashed, inner sep=2mm, outer sep=2mm, fill=green!50, fill opacity=0.2] at (rightp.center) {$\mathtt{X}$};
      \node[text=blue!10, draw=gray!50, thin, dashed, inner sep=2mm, outer sep=2mm, fill=blue!50, fill opacity=0.2] at (r3.center) {$\mathtt{X}$};
      \node[text=blue!10, draw=gray!50, thin, dashed, inner sep=2mm, outer sep=2mm, fill=blue!50, fill opacity=0.2] at ([shift={(0,0.3cm)}]rightr.center) {$\mathtt{X}$};
      \node[text=blue!10, draw=gray!50, thin, dashed, inner sep=2mm, outer sep=2mm, fill=blue!50, fill opacity=0.2] at ([shift={(0,-0.3cm)}]rightr.center) {$\mathtt{X}$};
      \node[text=blue!10, draw=gray!50, thin, dashed, inner sep=2mm, outer sep=2mm, fill=blue!50, fill opacity=0.2] at ([shift={(0,0.3cm)}]r2.center) {$\mathtt{X}$};
      \node[text=blue!10, draw=gray!50, thin, dashed, inner sep=2mm, outer sep=2mm, fill=blue!50, fill opacity=0.2] at ([shift={(0,-0.3cm)}]r2.center) {$\mathtt{X}$};
      \node[text=blue!10, draw=gray!50, thin, dashed, inner sep=2mm, outer sep=2mm, fill=blue!50, fill opacity=0.2] at ([shift={(0,0.5cm)}]Br.north) {$\mathtt{X}$};
      \node[text=blue!10, draw=gray!50, thin, dashed, inner sep=2mm, outer sep=2mm, fill=blue!50, fill opacity=0.2] at ([shift={(0,0.5cm)}]Cr.north) {$\mathtt{X}$};

    \end{pgfonlayer}
    \begin{pgfonlayer}{background}
      \node[draw, thin, dashed, inner sep=1mm, outer sep=2mm, fill=green!50, fill opacity=0.2, fit=(A) (B) (C)] (peq) {};

      \node[draw, thin, dashed, inner sep=1mm, outer sep=2mm, fill=orange!50, fill opacity=0.2, fit=(App) (Bpp) (Cpp)] (ppeq) {};

      \node[draw, thin, dashed, inner sep=1mm, outer sep=2mm, fill=orange!50, fill opacity=0.2, fit=(Api) (Bpi) (Cpi)] (pieq) {};

      \node[draw, thin, dashed, inner sep=1mm, outer sep=2mm, fill=orange!50, fill opacity=0.2, fit=(Aip)] (ipeq) {};

      \node[draw, thin, dashed, inner sep=1mm, outer sep=2mm, fill=orange!50, fill opacity=0.2, fit=(Aii)] (aiieq) {};
      \node[draw, thin, dashed, inner sep=1mm, outer sep=2mm, fill=orange!50, fill opacity=0.2, fit=(Bii)] (biieq) {};
      \node[draw, thin, dashed, inner sep=1mm, outer sep=2mm, fill=orange!50, fill opacity=0.2, fit=(Cii)] (ciieq) {};

      \node[draw, thin, dashed, inner sep=1mm, outer sep=2mm, fill=blue!50, fill opacity=0.2, fit=(Ar)] (areq) {};
      \node[draw, thin, dashed, inner sep=1mm, outer sep=2mm, fill=blue!50, fill opacity=0.2, fit=(Br)] (breq) {};
      \node[draw, thin, dashed, inner sep=1mm, outer sep=2mm, fill=blue!50, fill opacity=0.2, fit=(Cr)] (creq) {};
      
    \end{pgfonlayer}

    \begin{pgfonlayer}{main}
      \draw[->] (peq) to [out=180, in=180] node[draw, fill=white, pos=0.57] {$\link\kappa^+$ for $\stlc{\lambda}$} (ppeq);
      
      \draw[->] ([shift={(-0.1cm,0.2cm)}]areq.north) to [out=140, in=225] node[draw, fill=white, pos=0.57] {$\link\kappa^+$ for $\lamref{\lambda^{ref}}$} ([shift={(0,0.1cm)}]biieq.south); 

    \end{pgfonlayer}
  \end{tiny}
\end{tikzpicture}
\caption{Equivalence classes when giving different linking types to programs.}
\label{equivalence-classes}
\end{figure}

Reasoning about contextual equivalence means reasoning about the equivalence
classes that contain programs. Thus we can understand the effect of linking
types, and of the properties that guide them, by studying how the extensions
affect equivalence classes. In Figure~\ref{equivalence-classes},
we present three programs ($\gen{A}$, $\gen{B}$, and $\gen{C}$) valid
in both $\stlc\lambda$ and $\lamref{\lambda^{ref}}$. At the type
$\stlc{(int \rightarrow int) \rightarrow int}$, all three programs are
equivalent in $\stlc\lambda$, which we illustrate by putting
$\gen{A, B, C}$ in a single equivalence box. In $\stlc\lambda$, all
functions terminate, which means that calling the argument $\gen{f}$
zero, one, or two times before discarding the result is equivalent.
However, in $\lamref{\lambda^{ref}}$, $\gen{A}$, $\gen{B}$, and
$\gen{C}$ are all in different equivalence classes, since $\gen{f}$
may increment a counter, which means a context could detect the number
of times it was called.

The top of the diagram shows equivalence classes for
$\stlc\lambda^{\link\kappa}$/$\lamref{\lambda^{ref\link\kappa}}$. Here
we can see how equivalences can be changed by annotating these
functions with different linking types. Note that equivalence is only
defined at a given type, so we only consider when all three functions
have been given the same linking type. 

At the type
$\link{(int \rightarrow R^\circ int) \rightarrow R^\circ int}$ these
programs are all equivalent since this linking type requires that
$\gen{f}$ be pure. At the type
$\link{(int \rightarrow R^\bullet int) \rightarrow R^\bullet int}$
all three programs are in different equivalence classes, because the
linking type allows $\gen{f}$ to be impure, which could be used by a
context to distinguish the programs. At the type
$\link{(int \rightarrow R^\circ int) \rightarrow R^\bullet int}$ all
three programs are again equivalent. While the type allows the body to
be impure, since the argument $\gen{f}$ is pure, no difference can
be detected. The last linking type
$\link{(int \rightarrow R^\bullet int) \rightarrow R^\circ int}$ is a
type that can only be assigned to the program $\gen{A}$, because if
the argument $\gen{f}$ is impure but the result is pure the program
could not have called $\gen{f}$.

We can see here that $\link\kappa^+$ is the ``default'' embedding,
which has the important property that it preserves equivalence classes
from the original language. Notice that $\link\kappa^+$ for
$\stlc\lambda$ and $\lamref{\lambda^{ref}}$ both do this, and
send the respective source $\gen{(int \rightarrow int) \rightarrow int}$ to different types.

\section{Additional Applications of Linking Types}
\label{sec:examples}

This section contains examples of languages we would like to be able
to link with $\lamref{\lambda^{ref}}$ but which contain features that
require we either rule out such linking or give up on programmers
being able to reason in their source language (without linking
types).  We consider idealized languages here---and indeed, 
we believe that programmers would benefit from smaller, more 
special-purpose languages in a software project---but the ideas carry
through to full languages with different expressivity.

\subsection{Linearity in Libraries}
Substructural type systems are particularly useful for modeling
resources and for reasoning about where a resource must be used or
when consuming a resource should render it unusable to
others. Simple examples include network sockets and file handles, where 
opening creates the resource, reading consumes the resource and
possibly creates a new one, and closing consumes the resource.  An ML
programmer may want to use libraries written in a linear or affine
language (such as Rust) to ensure safe resource handling.  But if the 
language with linear or affine types allows values to cross the
linking boundary, ML needs to respect the linear or affine invariants
to ensure soundness.  For instance, if an ML component passes a value
as affine to a Rust component but retains a pointer to the value and
later tries to use it after it was consumed (in Rust), it violates the
affine invariant that every resource may be used at most once, making
the program crash.  Similarly, if an ML component never consumes a
linear value, it violates the linear invariant that every resouce must
be used exactly once, resulting in a resource leak. 

A fully abstract compiler would prevent linking in the above scenarios 
since two components that are equivalent in a linear/affine language
can easily be distinguished by a context that does not respect
linear/affine invariants. For instance, an affine function that
consumes its affine input and one that does not are equivalent if the
context cannot later try to consume the same input.  

We can use linking types to give non-linear languages access to
libraries with linear APIs.  Specifically, we would extend the types
of our non-linear source language $\lamref{\lambda^{ref}}$ as follows:

\vspace{-0.5cm}
\begin{small}
  \[
    \begin{array}{lcl}
     \link{\phi} & ::= & \link{unit \vo int \vo ref\, \tau \vo \tau \rightarrow \tau}\\
      \link{\tau} & ::= & \link{\phi \vo \phi^{L}}
    \end{array}
  \]
  \begin{minipage}[t]{0.5\textwidth}
    \vspace{-0.5cm}
    \[
      \begin{array}{lcl}
        \link{\kappa}^+(\lamref{unit}) & = & \link{unit} \\
        \link{\kappa}^+(\lamref{int}) & = & \link{int} \\
        \link{\kappa}^+(\lamref{ref\, \tau}) & = & \link{ref}\, \link{\kappa}^+(\lamref{\tau}) \\
        \link{\kappa}^+(\lamref{\tau_1 \rightarrow \tau_2}) & = & \link{\kappa}^+(\lamref{\tau_1}) \link{\rightarrow \kappa}^+(\lamref{\tau_2})\\
      \end{array}
    \]
  \end{minipage}
  \begin{minipage}[t]{0.5\textwidth}
    \vspace{-1.3cm}
    \[
      \begin{array}{lcl}
        \link{\kappa}^-(\link{\phi}) & = & \link{\kappa^L}^-(\link\phi) \\
        \link{\kappa}^-(\link{\phi^L}) & = & \link{\kappa^L}^-(\link\phi) \\
        \link{\kappa^L}^-(\link{unit}) & = & \lamref{unit} \\
        \link{\kappa^L}^-(\link{int}) & = & \lamref{int} \\
        \link{\kappa^L}^-(\link{ref\, \tau}) & = & \lamref{ref}\, \link{\kappa}^-(\link\tau) \\
        \link{\kappa^L}^-(\link{\tau_1 \rightarrow \tau_2}) & = & \link{\kappa}^-(\link{\tau_1}) \lamref\rightarrow \link{\kappa}^-(\link{\tau_2})\\ 
      \end{array}
    \]
  \end{minipage}
\end{small}

Note that the target of compilation would either need to
support linear types or enforce linearity at runtime---e.g., via
contracts {\`a} la Tov and Pucella \cite{tov10}. 

\subsection{Terminating Protocol Parsers}
For certain programming tasks, every program should terminate---for
instance, HTTP protocol parsing should never end up in an infinite
loop.  A programmer could implement such tasks using a special-purpose
language in which divergence is impossible.  We still, however, need
to link such terminating languages with general-purpose
languages---while the protocol parser should always terminate, the
server where it lives better not!   

A fully abstract compiler would have to prevent such linking, since two
components that are equivalent in a terminating language can easily be
distinguished by a context with nontermination.  For instance, a
function that calls its argument and discards the result, and one that
ignores its argument are equivalent if the context provides only
terminating functions as arguments, but not if the context provides a
function that diverges when called. 

We can use linking types to allow terminating and nonterminating
languages to interact.  Concretely, we can extend the types of our
nonterminating language $\lamref{\lambda^{ref}}$ with a terminating
function type, written $\link{\tau \rightarrow
  \tau{\downharpoonright}}$.  The extension is as follows, but we
elide cases of $\link{\kappa}^+$ and $\link{\kappa}^-$ that are the
same as in Figure~\ref{basic-linking-types}: 

\begin{small}

    \[
      \begin{array}{lcl}
        \link{\tau} & ::= & \link{unit \vo int \vo ref\, \tau \vo \tau \rightarrow \tau{\downharpoonright}\, \vo \tau \rightarrow \tau}
        \end{array}
      \]
    \begin{minipage}[t]{0.5\textwidth}
      \vspace{-.75cm}
      \[
        \begin{array}{lcl}      
          \link{\kappa}^+(\lamref{\tau_1 \rightarrow \tau_2}) & = & \link{\kappa}^+(\lamref{\tau_1}) \link{\rightarrow \kappa}^+(\lamref{\tau_2})\\
        \end{array}
      \]
    \end{minipage}
    \begin{minipage}[t]{0.5\textwidth}
      \vspace{-1.1cm}
      \[
        \begin{array}{lcl}
        \link{\kappa}^-(\link{\tau_1 \rightarrow \tau_2{\downharpoonright}}) & = & \link{\kappa}^-(\link{\tau_1}) \lamref\rightarrow \link{\kappa}^-(\link{\tau_2})\\
        \link{\kappa}^-(\link{\tau_1 \rightarrow \tau_2}) & = & \link{\kappa}^-(\link{\tau_1}) \lamref\rightarrow \link{\kappa}^-(\link{\tau_2})\\ 
      \end{array}
    \]
  \end{minipage}
\end{small}

The typing rules (elided) would likely need to rely on some syntactic
termination check for functions ascribed the terminating arrow type.  We
could also imagine making the terminating arrow rely on a runtime 
timeout.  The latter would require a new application typing rule to
reflect that sometimes applying a terminating function might return a
nonce value indicating that computation was cut off, and our language
would need to be trivially extended with sum types to handle that
possibility in programs.  

\subsection{Surfacing Cost of Computation}
Some security vulnerabilities rely on the fact that the cost of a
computation may be discernable (e.g., by observing time, or CPU or memory
consumption).  To prove the absence of such vulnerabilities, we could
remove the mechanism of observation---but this is likely impossible,
since even if we remove timing from our language, if the program
communicates over the network timing can happen on other systems.
A more promising strategy is to 
% Proving the absence of such vulnerabilities requires impossible, since
% even if we remove timing from our language, if the program
% communicates over the network timing can happen on other systems.  A
% better alternative
% either removing the mechanism of observation---which is likely
% impossible, since even if we remove timing from our language, if the
% program communicates over the network timing can happen on other
% systems---or 
introduce the notion of cost (time or space) into the model and then
prove that various branches are indistinguishable in that model (see,
e.g. \cite{blelloch13}, \cite{hoffmann12}, \cite{cicek17}).
Nonetheless, one would not want to have to write
non-security-sensitive parts of programs in one of these cost-aware
languages.  This motivates a linking-types extension of a
non-cost-aware language---in this case again our idealized 
$\lamref{\lambda^{ref}}$---with a notion of computations with cost.  As
before, we only show differences from Figure~\ref{basic-linking-types}: 

\begin{small}
  \[
      \link{\tau} ::= \link{unit \vo int \vo ref\, \tau \vo \tau \rightarrow C^\bullet \tau \vo \tau \rightarrow C^N \tau}
  \]
  \begin{minipage}[t]{0.5\textwidth}
    \vspace{-0.75cm}
    \[
      \begin{array}{lcl}
        \link{\kappa}^+(\lamref{\tau_1 \rightarrow \tau_2}) & = & \link{\kappa}^+(\lamref{\tau_1}) \link{\rightarrow C^\bullet \kappa}^+(\lamref{\tau_2})\\
      \end{array}
    \]
  \end{minipage}
  \begin{minipage}[t]{0.5\textwidth}
    \vspace{-1.1cm}
    \[
      \begin{array}{lcl}
        \link{\kappa}^-(\link{\tau_1 \rightarrow C^\bullet \tau_2}) & = & \link{\kappa}^-(\link{\tau_1}) \lamref\rightarrow \link{\kappa}^-(\link{\tau_2})\\
        \link{\kappa}^-(\link{\tau_1 \rightarrow C^N \tau_2}) & = & \link{\kappa}^-(\link{\tau_1}) \lamref\rightarrow \link{\kappa}^-(\link{\tau_2})\\

      \end{array}
    \]

  \end{minipage}
\end{small}

The type system is modal---computations $\link{C^N\tau}$ have a known
cost $\link{N}$, and computations $\link{C^\bullet\tau}$ have an unknown
cost. Fully abstract compilation from a cost-aware language and the
above extended language would only allow known-cost computations to be
passed to the cost-aware language. As before, this relies upon the
target language supporting a type system that is at least as
expressive, such that it can safely separate the known-cost and
unknown-cost modalities. 

This application of linking types echos the work by
D'Silva~\etal~\cite{dsilva15}, which discusses enriching the model in
which properties are investigated to encompass side channels like 
timing.  While their work investigates machine models, ours relies
upon type systems in the language where linking takes place. Further,
D'Silva~\etal~envision programmers would \emph{opt in} to security 
properties via annotations that would change how the compiler treated
a piece of code, whereas we envision that the compiler would preserve
source equivalences by default and programmers would have to \emph{opt
  out} of the default fully abstract compilation by 
using linking types.  We believe that our approach can be used with
other side channels as well, provided sufficient mechanisms exist to
distinguish computations that might reveal information from those that
cannot. 

\subsection{Gradual Typing}
As we have already shown, linking types are useful when linking more precisely
and less precisely typed languages. Taken to an extreme, we can add linking
types to a un(i)typed language to facilitate sound linking with a statically
typed language.  We can do this by starting with a language with a single type, the
dynamic type, and then constructing an extension that adds further
types. A typed target language would then allow code compiled from a
different, typed, source language to be linked with this gradually
typed language. A fully abstract compiler for the extended language
would have to make use of run-time checks at the boundaries between
typed and untyped code, analogous to sound gradual typing.
\newcommand\cparagraph{\noindent\underline{Discussion}}
\newcommand\citem{~$\bullet$~}
\newcommand{\cday}[1]{\vspace{0.2cm}\noindent\textsf{\textbf{{#1}}}~~}

\section{Bringing Linking Types to Your Language}

To understand linking types and the way they interact with existing
languages, we consider an example of how a language designer would
incorporate them and discuss their usefulness and
viability ({\`a} la Cardelli~\cite{cardelli97:linking}).

\cday{Day 1: Fully abstract compiler} As a first step, the language designer
implements and proves fully abstract a type-directed compiler for her language
$\mathcal{A}$. To make it more concrete, you can consider $\mathcal{A}$ to be
the language $\stlc{\lambda}$ from earlier in the paper, but this scenario is
general --- you could equally consider $\mathcal{A}$ to be a language like
OCaml. The compiler targets a typed low-level intermediate language
$\mathcal{L}$, using an appropriate type translation to guarantee that
equivalences are preserved. This, concretely, could be a target like
$\target{\lambda^{ref}_{exn}}$, but could also be a richly-typed version of
LLVM. All linking should occur in $\mathcal{L}$, which means the subsequent
passes, to LLVM, assembly, or another target, need not be fully abstract.

\cparagraph
\citem Full abstraction is a key part of linking types, as it is
required to preserve the equivalences that programmers rely upon for
reasoning.
\citem The representative terms added to the linking-types-extended
language are used in the proof of full abstraction, which essentially
requires showing that target contexts can be back-translated to
equivalent source contexts.  
\citem While we use static types in our target to ensure full abstraction---
and gain tooling benefits from it (explored in Day 3)---we can also use
dynamic checks when appropriate
(e.g.~\cite{patrignani15,devriese16}).
\citem We are currently designing a language like $\mathcal{L}$, which we expect
to be similar to a much more richly typed version of LLVM, such that types could
be erased and existing LLVM code-generation infrastructure could be used (as
discussed by Ahmed at SNAPL'15~\cite{ahmed15:snapl}).

\cday{Day 2: Linking with more expressive code}
$\mathcal{A}$ programmers are happy using the above compiler since they can
reason in terms of $\mathcal{A}$ semantics, even when using libraries
directly implemented in $\mathcal{L}$ or compiled from other
languages. 
But, soon the language designer's users ask to link
their code with a $\mathcal{B}$ language library with features in
$\mathcal{L}$ but not in $\mathcal{A}$, something that the fully
abstract compiler currently prevents. In the example used earlier in the paper,
$\mathcal{B}$ would be $\lamref{\lambda^{ref}}$, and the additional feature
would be mutable references, but again, this is a general process that could
apply to other features.

The compiler writer introduces a linking-types extension to capture
the inexpressible features for her $\mathcal{A}$ programmers. She
implements a type checker for the fully elaborated linking types and
extends her fully abstract compiler to handle the extended
$\mathcal{A^{\link\kappa}}$ types.

\cparagraph
\citem While the linking types will in general be a new type system,
no impact is seen on type inference, because linking types are never
inferred: first the program will have source types inferred, and then
all source types will be lifted to the linking types, using the
programmer-specified annotations where present and the default
$\link\kappa^+$ embedding where annotations are absent.

\cday{Day 3: When can components in two languages be linked?} Happily able to
link with other languages, the $\mathcal{A}$ programmer uses the tooling
associated with the $\mathcal{A}$ and $\mathcal{B}$ compilers to determine when
a $\mathcal{B}$ component can be used at a linking point. The tool uses the
compiler to translate the $\mathcal{B}$ component's type $\tau_{\mathcal{B}}$ to
an $\mathcal{L}$ type $\tau_{\mathcal{L}}$ and then attempts to back-translate
$\tau_{\mathcal{L}}$ to an $\mathcal{A}$ type $\tau_{\mathcal{A}}$ by inverting
the $\mathcal{A}$ compiler's type translation. Should this succeed, the
component can be used at the type $\tau_{\mathcal{A}}$. This functionality
allows the programmer to easily work on components in both $\mathcal{A}$ and
$\mathcal{B}$ at once while getting \emph{cross-language type errors} if the
interfaces do not match. In the example used earlier in the paper, such a type
error showed up when trying to link the $\lamref{\lambda^{ref}}$ counter library
with the $\stlc\lambda$ client that had not been annotated.

\cparagraph
\citem This functionality depends critically on the type-directed
nature of our compilers and the presence of types in the low-level
intermediate language $\mathcal{L}$, where the types become the
\emph{medium} through which we can provide useful static feedback to
the programmer. 
\citem While linking these components together relies upon shared
calling conventions, this is true of any linking. Currently,
cross-language linking often relies upon C calling conventions.

\cday{Day 4: Backwards compatibility for programmers}
At the same time, another programmer continues to use $\mathcal{A}$,
unaware of the $\mathcal{A^{\link\kappa}}$ extension introduced in Day 2, since linking
types are \emph{optional} annotations. At lunch, she learns about
linking types and realizes that the $\mathcal{C}$ language she uses
could benefit from the $\mathcal{L}$ linking ecosystem. She asks the compiler writer for a
$\mathcal{C}$-to-$\mathcal{L}$ compiler.

\cparagraph
\citem Linking types are entirely opt-in---a programmer can use a
language that has been extended with them and benefit from the
compiler tool-chain without knowing anything about them. Only when
she wants to link with code that could violate her source-level
reasoning does she need to deal with linking types.
\citem FFIs are usually considered ``advanced material'' in language
documentation primarily due to the difficulty of using them safely.  
Since linking types enable safe cross-language linking, we hope
that linking-type FFIs will not be considered such an advanced
topic. 

\cday{Day 5: Backwards compatibility for language designers}
Never a dull day for the compiler writer: she starts implementing a
fully abstract compiler from $\mathcal{C}$ to $\mathcal{L}$, but
realizes that $\mathcal{L}$ is not rich enough to capture the
properties needed. She extends $\mathcal{L}$ to $\mathcal{L^*}$, and
proves fully abstract the translation from $\mathcal{L}$ to
$\mathcal{L^*}$. Since full abstraction proofs compose, this means
that she immediately has a fully abstract compiler from
$\mathcal{A^{\link\kappa}}$ to $\mathcal{L^*}$. She then implements
a fully abstract compiler from $\mathcal{C}$ to $\mathcal{L^*}$.
Programmers can then link $\mathcal{A^{\link\kappa}}$ components and $\mathcal{C}$
components provided that the former do not use $\mathcal{C}$ features
that cannot be expressed in $\mathcal{A^{\link\kappa}}$.  Luckily for
the compiler writer, the proofs mean that the behavior of $\mathcal{A^{\link\kappa}}$, even in the presence
of linking, was fully specified before and remains so.  Hence, $\mathcal{A}$ and
$\mathcal{A^{\link\kappa}}$ programmers need not even know about the change from
$\mathcal{L}$ to $\mathcal{L^*}$. 

\cparagraph
\citem While implementing fully abstract compilers is nontrivial, 
the linking-types strategy permits a gradual evolution, not requiring
redundant re-implementation and re-proof whenever changes to the target
language are made.
\citem More generally, the proofs of full abstraction mean that the
compiler and the target are irrelevant for programmers---behavior is
entirely specified at the level of the (possibly extended by linking
types) source.

\section{Research Plan and Challenges}

We are currently studying the use of linking types to facilitate
building multi-language programs that may consist of components
from the following: an idealized ML
(essentially System F with references); a simple linear language; a 
language with first-class control; and a terminating language.  We
plan to develop a richly typed target based on Levy's
call-by-push-value (CBPV)~\cite{levy01:phd} that can support fully
abstract compilation from our linking-types-extended languages.
Zdancewic (personal communication on Vellvm2) has recently demonstrated a machine
equivalence between a variant of CBPV and an LLVM-like SSA-based IR so
this provides a path from our current intended target to a richly typed LLVM.

One critical aspect of such a type system is that it should be able to identify
when a component is free of a given effect, even though the component may use
that effect internally. For instance, a component that throws exceptions
internally but handles them all should be assigned an exception-free type. We expect
to draw inspiration from the effect-masking in the Koka language~\cite{leijen14}, where
mutable references that never escape do not cause a computation to be marked as
effectful.

Realizing such a multi-language programming platform involves a number
of challenges.  First, implementing fully abstract compilers is
nontrivial, though there has been significant recent progress by both
our group and others that we expect to draw
upon~\cite{ahmed08:tccpoe,ahmed11:epcps,devriese16,bowman15:nifree,new16:facue,patrignani15}.
Second, low-level languages such as LLVM and assembly are typically
non-compositional which makes it hard to support high-level
compositional reasoning.  In recent work, we have designed a
compositional typed assembly language that we think offers a blueprint
for designing other low-level typed
IRs~\cite{patterson17:funtal,patterson17:funtal-tr}.  Finally, we have
only begun investigating how to combine different linking-types
extensions. The linking-types extensions we are considering are based
on type-and-effect systems, so we believe we can create a lattice of
these extensions analogous to an effect lattice.

\section{Conclusion}

Large software systems are written using combinations of many languages. But
while some languages provide powerful tools for reasoning \emph{in} the
language, none support reasoning \emph{across} multiple languages. Indeed, the
abstractions that languages purport to present do not actually cohere because
they do not allow the programmer to reason solely about the code she writes.
Instead, the programmer is forced to think about the details of particular
compilers and low-level implementations, and to reason about the target code
that her compiler generates.

With \emph{linking types}, we propose that language designers incorporate
linking into their language designs and provide programmers a means to specify
linking with behavior and types inexpressible in their language. There are many
challenges in how to design linking types, depending on what features exist in
the languages, but only through accepting this challenge can we reach what has
long been promised---an ecosystem of languages, each suited to a particular task
yet stitched together seamlessly into a single large software project.

\subparagraph*{Acknowledgements} The authors are grateful to Matthias Felleisen
for valuable discussion and feedback on linking types. This research is
supported in part by the NSF (grants CCF-1453796 and CCF-1422133) and a Google
Faculty Research Award.

\bibliography{Patterson}

\begin{thebibliography}{10}

\bibitem{ahmed15:snapl}
Amal Ahmed.
\newblock {Verified Compilers for a Multi-Language World}.
\newblock In Thomas Ball, Rastislav Bodik, Shriram Krishnamurthi, Benjamin~S.
  Lerner, and Greg Morrisett, editors, {\em 1st Summit on Advances in
  Programming Languages (SNAPL 2015)}, volume~32 of {\em Leibniz International
  Proceedings in Informatics (LIPIcs)}, pages 15--31, 2015.

\bibitem{ahmed08:tccpoe}
Amal Ahmed and Matthias Blume.
\newblock Typed closure conversion preserves observational equivalence.
\newblock In {\em {I}nternational {C}onference on {F}unctional {P}rogramming
  ({ICFP}), Victoria, British Columbia, Canada}, pages 157--168, September
  2008.

\bibitem{ahmed11:epcps}
Amal Ahmed and Matthias Blume.
\newblock An equivalence-preserving {CPS} translation via multi-language
  semantics.
\newblock In {\em {I}nternational {C}onference on {F}unctional {P}rogramming
  ({ICFP}), Tokyo, Japan}, pages 431--444, September 2011.

\bibitem{benton09}
Nick Benton and Chung-Kil Hur.
\newblock Biorthogonality, step-indexing and compiler correctness.
\newblock In {\em {I}nternational {C}onference on {F}unctional {P}rogramming
  ({ICFP}), Edinburgh, Scotland}, September 2009.

\bibitem{benton10}
Nick Benton and Chung-Kil Hur.
\newblock Realizability and compositional compiler correctness for a
  polymorphic language.
\newblock Technical Report MSR-TR-2010-62, Microsoft Research, April 2010.

\bibitem{blelloch13}
Guy~E. Blelloch and Robert Harper.
\newblock Cache and {I/O} efficient functional algorithms.
\newblock In {\em {ACM} {S}ymposium on {P}rinciples of {P}rogramming
  {L}anguages ({POPL}), Rome, Italy}, pages 39--50, January 2013.

\bibitem{bowman15:nifree}
William~J. Bowman and Amal Ahmed.
\newblock Noninterference for free.
\newblock In {\em {I}nternational {C}onference on {F}unctional {P}rogramming
  ({ICFP}), Vancouver, British Columbia, Canada}, September 2015.

\bibitem{cardelli97:linking}
Luca Cardelli.
\newblock Program fragments, linking, and modularization.
\newblock In {\em {ACM} {S}ymposium on {P}rinciples of {P}rogramming
  {L}anguages ({POPL}), Paris, France}, pages 266--277, January 1997.

\bibitem{cicek17}
Ezgi {\c C}i{\c c}ek, Gilles Barthe, Marco Gaboardi, Deepak Garg, and Jan
  Hoffmann.
\newblock Relational cost analysis.
\newblock In {\em {ACM} {S}ymposium on {P}rinciples of {P}rogramming
  {L}anguages ({POPL}), Paris, France}, January 2017.

\bibitem{devriese16}
Dominique Devriese, Marco Patrignani, and Frank Piessens.
\newblock Fully-abstract compilation by approximate back-translation.
\newblock In {\em {ACM} {S}ymposium on {P}rinciples of {P}rogramming
  {L}anguages ({POPL}), St. Petersburg, Florida}, 2016.

\bibitem{dsilva15}
Vijay D'Silva, Mathias Payer, and Dawn Song.
\newblock The correctness-security gap in compiler optmization.
\newblock In {\em Language-theoretic Security IEEE Security and Privacy
  Workshop (LangSec)}, 2015.

\bibitem{felleisen90}
Matthias Felleisen.
\newblock On the expressive power of programming languages.
\newblock In {\em Science of Computer Programming}, pages 134--151.
  Springer-Verlag, 1990.

\bibitem{gu15:deepspec}
Ronghui Gu, J{\'e}r{\'e}mie Koenig, Tahina Ramananandro, Zhong Shao,
  Xiongnan~(Newman) Wu, Shu-Chun Weng, Haozhong Zhang, and Yu~Guo.
\newblock Deep specifications and certified abstraction layers.
\newblock In {\em {ACM} {S}ymposium on {P}rinciples of {P}rogramming
  {L}anguages ({POPL}), Mumbai, India}, pages 595--608, January 2015.

\bibitem{hoffmann12}
Jan Hoffmann, Klaus Aehlig, and Martin Hofmann.
\newblock {Resource Aware ML}.
\newblock In {\em {24rd International Conference on Computer Aided Verification
  (CAV'12)}}, volume 7358 of {\em Lecture Notes in Computer Science}, pages
  781--786. Springer, 2012.

\bibitem{hur11}
Chung-Kil Hur and Derek Dreyer.
\newblock A {K}ripke logical relation between {ML} and assembly.
\newblock In {\em {ACM} {S}ymposium on {P}rinciples of {P}rogramming
  {L}anguages ({POPL}), Austin, Texas}, January 2011.

\bibitem{kang2016lightweight}
Jeehoon Kang, Yoonseung Kim, Chung-Kil Hur, Derek Dreyer, and Viktor Vafeiadis.
\newblock Lightweight verification of separate compilation.
\newblock In {\em {ACM} {S}ymposium on {P}rinciples of {P}rogramming
  {L}anguages ({POPL}), St. Petersburg, Florida}, pages 178--190. ACM, 2016.

\bibitem{kumar14}
Ramana Kumar, Magnus~O. Myreen, Michael Norrish, and Scott Owens.
\newblock {CakeML} : A verified implementation of {ML}.
\newblock In {\em {ACM} {S}ymposium on {P}rinciples of {P}rogramming
  {L}anguages ({POPL}), San Diego, California}, January 2014.

\bibitem{leijen14}
Daan Leijen.
\newblock Koka: Programming with row polymorphic effect types.
\newblock In {\em {M}athematically {S}tructured {F}unctional {P}rogramming,
  Grenoble, France}, April 2014.

\bibitem{leroy06}
Xavier Leroy.
\newblock Formal certification of a compiler back-end or: programming a
  compiler with a proof assistant.
\newblock In {\em {ACM} {S}ymposium on {P}rinciples of {P}rogramming
  {L}anguages ({POPL}), Charleston, South Carolina}, January 2006.

\bibitem{leroy09:jar}
Xavier Leroy.
\newblock A formally verified compiler back-end.
\newblock {\em Journal of Automated Reasoning}, 43(4):363--446, 2009.

\bibitem{levy01:phd}
Paul~Blain Levy.
\newblock {\em Call-by-Push-Value}.
\newblock {Ph.~D.} dissertation, Queen Mary, University of London, London, UK,
  March 2001.

\bibitem{lochbihler10}
Andreas Lochbihler.
\newblock Verifying a compiler for {J}ava threads.
\newblock In {\em European Symposium on Programming (ESOP)}, March 2010.

\bibitem{neis15}
Georg Neis, Chung-Kil Hur, Jan-Oliver Kaiser, Craig McLaughlin, Derek Dreyer,
  and Viktor Vafeiadis.
\newblock Pilsner: A compositionally verified compiler for a higher-order
  imperative language.
\newblock In {\em {I}nternational {C}onference on {F}unctional {P}rogramming
  ({ICFP}), Vancouver, British Columbia, Canada}, August 2015.

\bibitem{new16:facue}
Max~S. New, William~J. Bowman, and Amal Ahmed.
\newblock Fully abstract compilation via universal embedding.
\newblock In {\em {I}nternational {C}onference on {F}unctional {P}rogramming
  ({ICFP}), Nara, Japan}, September 2016.

\bibitem{patrignani15}
Marco Patrignani, Pieter Agten, Raoul Strackx, Bart Jacobs, Dave Clarke, and
  Frank Piessens.
\newblock Secure compilation to protected module architectures.
\newblock {\em ACM Transactions on Programming Languages and Systems},
  37(2):6:1--6:50, April 2015.

\bibitem{patterson17:funtal}
Daniel Patterson, Jamie Perconti, Christos Dimoulas, and Amal Ahmed.
\newblock {FunTAL}: Reasonably mixing a functional language with assembly.
\newblock In {\em {ACM SIGPLAN Conference on Programming Language Design and
  Implementation (PLDI)}, Barcelona, Spain}, June 2017.
\newblock To appear. Available at
  \texttt{http://www.ccs.neu.edu/\linebreak[0]home/\linebreak[0]amal/\linebreak[0]papers/\linebreak[0]funtal.pdf}.

\bibitem{patterson17:funtal-tr}
Daniel Patterson, Jamie Perconti, Christos Dimoulas, and Amal Ahmed.
\newblock {FunTAL}: Reasonably mixing a functional language with assembly
  (technical appendix).
\newblock Available at
  \texttt{http://www.ccs.neu.edu/\linebreak[0]home/\linebreak[0]amal/\linebreak[0]papers/\linebreak[0]funtal-tr.pdf},
  April 2017.

\bibitem{perconti14:fca}
James~T. Perconti and Amal Ahmed.
\newblock Verifying an open compiler using multi-language semantics.
\newblock In {\em European Symposium on Programming (ESOP)}, April 2014.

\bibitem{sevcik11}
Jaroslav Sevcik, Viktor Vafeiadis, Francesco~Zappa Nardelli, Suresh
  Jagannathan, and Peter Sewell.
\newblock Relaxed-memory concurrency and verified compilation.
\newblock In {\em {ACM} {S}ymposium on {P}rinciples of {P}rogramming
  {L}anguages ({POPL}), Austin, Texas}, 2011.

\bibitem{stewart15}
Gordon Stewart, Lennart Beringer, Santiago Cuellar, and Andrew~W. Appel.
\newblock Compositional compcert.
\newblock In {\em {ACM} {S}ymposium on {P}rinciples of {P}rogramming
  {L}anguages ({POPL}), Mumbai, India}, 2015.

\bibitem{tov10}
Jesse Tov and Riccardo Pucella.
\newblock Stateful contracts for affine types.
\newblock In {\em European Symposium on Programming (ESOP)}, March 2010.

\bibitem{zhao13:ssa}
Jianzhou Zhao, Santosh Nagarakatte, Milo M.~K. Martin, and Steve Zdancewic.
\newblock Formal verification of {SSA}-based optimizations for {LLVM}.
\newblock In {\em {ACM SIGPLAN Conference on Programming Language Design and
  Implementation (PLDI)}, Seattle, Washington}, June 2013.

\end{thebibliography}

\end{document}